\documentclass[12pt]{article}

\usepackage{scicite}
\usepackage{graphicx}
\usepackage{times}
\usepackage{xr}
\makeatletter
\newcommand*{\addFileDependency}[1]{
  \typeout{(#1)}
  \@addtofilelist{#1}
  \IfFileExists{#1}{}{\typeout{No file #1.}}
}
\makeatother

\newcommand*{\myexternaldocument}[1]{
    \externaldocument{#1}
    \addFileDependency{#1.tex}
    \addFileDependency{#1.aux}
}
\myexternaldocument{supp}

\newenvironment{sciabstract}{
\begin{quote} \bf}
{\end{quote}}

\title{Subgap spectroscopy along hybrid nanowires by nm-thick tunnel barriers} 

\author
{Vukan Levajac$^{1 \ast}$, Ji-Yin Wang$^{1,2 \ast \dag}$, Grzegorz P. Mazur$^1$, Cristina Sfiligoj$^1$,\\ Mathilde Lemang$^1$, Jan Cornelis Wolff$^1$, Alberto Bordin$^1$, Ghada Badawy$^3$,\\ Sasa Gazibegovic$^3$, Erik P. A. M. Bakkers$^3$, Leo P. Kouwenhoven$^{1}$\\
\\
\normalsize{$^{1}$QuTech and Kavli Institute of Nanoscience, Delft University of Technology,}\\
\normalsize{2628GA Delft, The Netherlands}\\
\normalsize{$^{2}$Beijing Academy of Quantum Information Sciences,}\\
\normalsize{100193 Beijing, China}\\
\normalsize{$^{2}$Department of Applied Physics, Eindhoven University of  Technology,}\\
\normalsize{5600MB Eindhoven, The Netherlands}\\
\normalsize{$^\ast$These authors contributed equally to this work.}\\ 
\normalsize{$^\dag$To whom correspondence should be addressed; E-mail:}\\ 
\normalsize{wangjiyinshu@gmail.com.}
}

\date{}

\begin{document} 
\baselineskip24pt
\maketitle 

\begin{sciabstract}
  Tunneling spectroscopy is widely used to examine the subgap spectra in semiconductor-superconductor nanostructures when searching for Majorana zero modes (MZMs). Typically, semiconductor sections controlled by local gates at the ends of hybrids serve as tunnel barriers. Besides detecting states only at the hybrid ends, such gate-defined tunnel probes can cause the formation of non-topological subgap states that mimic MZMs. Here, we develop an alternative type of tunnel probes to overcome these limitations. After the growth of an InSb-Al hybrid nanowire, a precisely controlled in-situ oxidation of the Al shell is performed to yield a nm-thick Al oxide layer. In such thin isolating layer, tunnel probes can be arbitrarily defined at any position along the hybrid nanowire by shadow-wall angle-deposition of metallic leads. This allows us to make multiple tunnel probes along single nanowire hybrids and to successfully identify Andreev bound states (ABSs) of various spatial extension residing along the hybrids. 
\end{sciabstract}

\section*{Introduction}

Topological superconductors have received significant attention in the condensed matter physics community over the last decade due to their potential application in fault-tolerant quantum computation \cite{kitaev_physusp_2001, kitaev_annalsofphys2003, nayak_revmodphys2008, dassarma_npjquantum2015}. In III-V semiconducting nanowires with thin superconducting shells a topological phase transition is predicted to occur at a sufficiently high magnetic field \cite{oreg_prl_2010,lutchyn_prl_2010}. An essential precondition for this is a hybridization mechanism in which the superconductivity is induced in the semiconducting nanowire with tunable chemical potential, strong spin-orbit interaction and large $g$ factor. The sophisticated interplay of these physical phenomena has motivated in-depth theoretical studies and state-of-the-art material developments \cite{aguado_nuovo_cim_2017,lutchyn_natrevmatt_2018,prada_natrevphys_2020}– with a goal of reaching topological superconducting phase in hybrid nanowires. Hallmarks of the topologically non-trivial phase are Majorana zero modes (MZMs) - zero energy hybrid modes localized at two ends of a hybrid nanowire.

Tunneling spectroscopy is commonly used to investigate the energy spectrum in hybrid nanowires and search for MZMs by examining the presence of zero energy states at nanowire ends. In such experiments, a normal metal lead is tunnel-coupled to an end of a hybrid nanowire and serves as a tunnel probe. The tunneling conductance is measured as a function of an applied bias voltage between the tunnel probe and the superconducting shell on the nanowire. Zero bias peaks (ZBPs) measured in such spectroscopies at hybrid nanowire ends indicate the presence of zero energy end states and were the first reported signatures of MZMs in hybrid nanowires \cite{mourik_science_2012,das_natphys_2012,deng_nanolett_2012}. A semiconducting nanowire section where the superconducting shell ends is generally used to create a tunnel barrier and a local tunnel gate is needed to define and control the barrier profile. Advanced numerical modellings of realistic devices have shown that low energy states can be localized at the end of a hybrid nanowire due to smooth variations in the electrostatic potential induced by the tunnel gate \cite{prada_prb_2012, vuik_scipostphys_2019,avila_commphys_2019}. A recent study on three-terminal hybrid nanowire devices has reported such zero energy states of trivial origin coincidentally appearing at both nanowire ends and mimicking the end-to-end correlation of MZMs \cite{wang_prb_2022}. Therefore, due to the smooth potential effects, ambiguous signatures of MZMs can be measured by tunnel probes with semiconducting tunnel barriers \cite{pan_prr_2020}. Another limitation of these tunnel probes is that the tunneling spectroscopy is performed only at the ends of a hybrid nanowire. Therefore, a reopening of an induced gap in the hybrid bulk at the topological phase transition can only be detected in nonlocal conductance measurements on three-terminal hybrid nanowire devices \cite{rosdahl_prb_2018}. Measuring the hybrid bulk directly in the local tunneling spectroscopy is additionally motivated by recent theoretical studies showing that disorder in a hybrid nanowire can result in MZMs being localized inside the hybrid bulk and undetectable at its ends \cite{ahn_prmater_2021,woods_prapplied_2021}. An experimental work has shown the possibility to use Al oxide as a tunnel barrier for hybrid nanowires with superconducting Al \cite{grivnin_natcomm_2019}. However, the lack of in-situ fabrication of the Al oxide and physical etching on the oxide layer during fabrication flow lead to a low-quality tunnel barriers - causing a quite softened superconducting gap.

Here, we develop a new type of tunnel barriers for tunneling spectroscopy in hybrid nanowires in order to overcome the limitations set by the semiconducting tunnel barriers. We fabricate InSb-Al hybrid nanowires in which a nm-thick dielectric layer of Al oxide covers the hybrid and can be used to tunnel couple it to a normal metal lead. In comparison with the reference \cite{grivnin_natcomm_2019}, the Al oxide layer is fabricated in-situ, which improves the quality of the tunnel probes in our work. Such tunnel probes have a sharp barrier profile set by the thickness and the band offset of the Al oxide layer and therefore do not cause smooth variations of electrostatic potential. In addition, the Al oxide layer extends over the entire length of the hybrid and allows for a formation of tunnel probes at any position along the nanowire. We exploit these advantages and fabricate multiple tunnel probes along single hybrid nanowires in order to investigate the longitudinal evolution of their energy spectra. By comparing the tunneling spectroscopy results obtained at different positions along the same nanowire, Andreev bound states (ABSs) of various spatial extension can be identified at the end and inside the bulk of the hybrids.

\section*{Results}

\subsection*{Device design}

\begin{figure}[!t] 
\centering
\includegraphics[width=\linewidth] {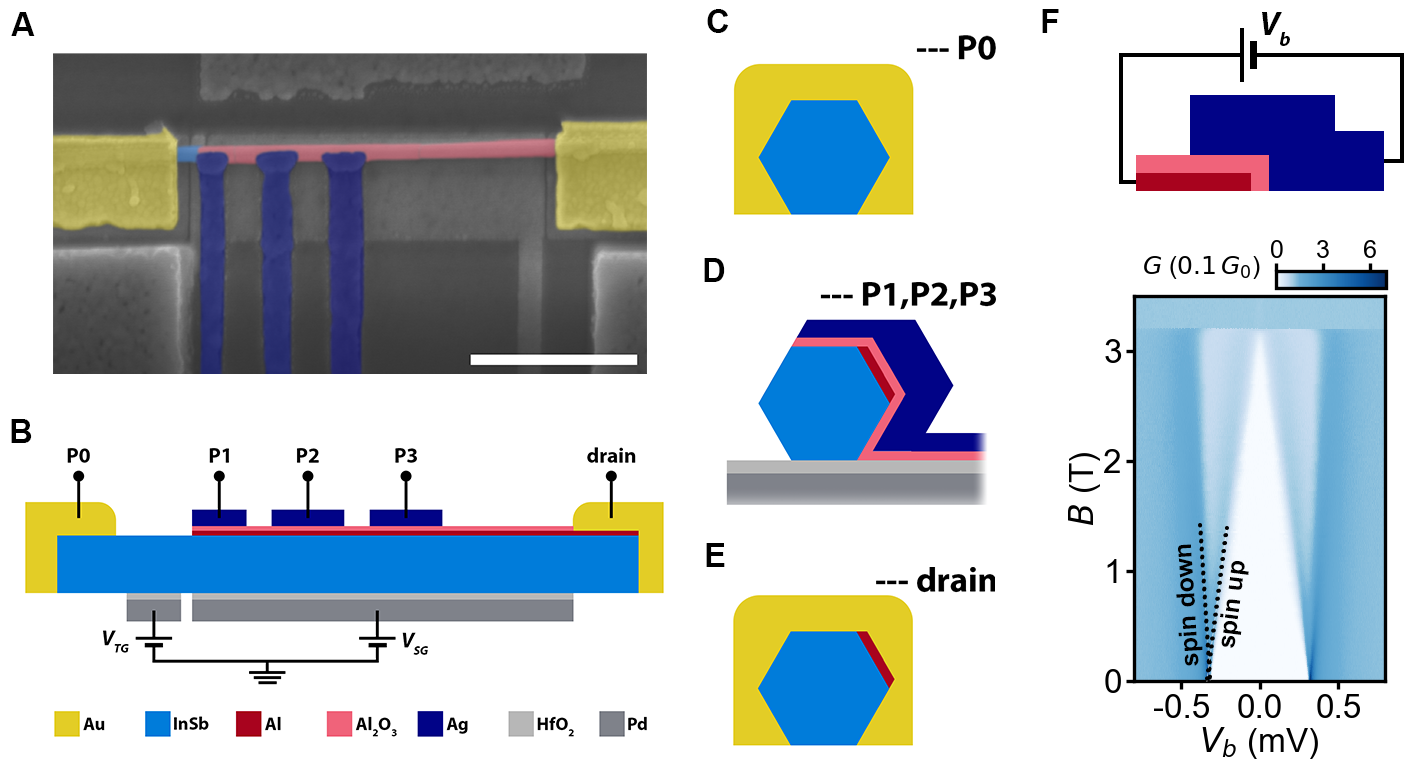}
\caption{\textbf{Hybrid nanowire devices with nm-thick tunnel barriers}: (\textbf{A}) False-colored SEM image of a representative device. A nm-thick layer of $\mathrm{Al_2O_3}$ (pink) fully covers the Al (red) shell that is visible in the schematical cross-sections (B to E). Three Ag (navy) leads are defined on top of the Al oxide layer along the hybrid. Two Au (yellow) leads contact the semiconducting InSb (light blue) nanowire on the left and the hybrid on the right. The white scale bar corresponds to $1\,\mathrm{\mu m}$.  (\textbf{B}) A schematical longitudinal cross-section along the device with two Pd (dark grey) gates coupled to the nanowire via dielectric $\mathrm{HfO_2}$ (light grey). Voltages $V_{TG}$ and $V_{SG}$ are applied to the tunnel gate and the super gate, respectively. Four probes P0, P1, P2 and P3 are tunnel-coupled to the hybrid nanowire contacted by the right drain lead. The probe P0 utilizes the semiconducting tunnel barrier and the probes P1, P2 and P3 use nm-thick tunnel barriers in the Al oxide layer. (\textbf{C} to \textbf{E}) Schematical transversal cross-sections through the tunnel probes and the drain. (\textbf{F}) A schematical perpendicular cross-section of a planar tunnel junction with an Al oxide layer as the tunnel barrier between an Al and an Ag film as the leads (top). Tunneling conductance $G$ of the junction as a function of a bias voltage $V_b$ and an in-plane magnetic field $B$ (bottom). A superconducting gap of $325 \pm 5\,\mathrm{\mu eV}$ and a critical in-plane field of $\sim 3.3\,\mathrm{T}$ can be extracted for the Al film from the tunneling spectroscopy. The dimensions of all the schematics are not scaled. }\label{fig:1}
\end{figure}

Hybrid nanowires that utilize nm-thick tunnel barriers are introduced in Fig. \ref{fig:1}. A false-colored scanning-electron microscopy (SEM) image of a representative device is shown in Fig. \ref{fig:1}A and a schematical longitudinal cross-section along the device is displayed in Fig. \ref{fig:1}B. A superconducting Al (red) film is grown by the shadow-wall lithography technique \cite{heedt_natcomm_2021,borsoi_afm_2021} on a semiconducting InSb (light blue) nanowire \cite{badawy_nanolett_2019}. By a subsequent in-situ oxidation, the Al film is partially oxidized to form a dielectric Al oxide - $Al_2O_3$ (pink) layer that covers the hybrid. The shadow-wall lithography technique is used to define three normal Ag (navy) leads along the nanowire on top of the Al oxide layer. Two Au (yellow) leads contact the bare semiconducting nanowire part on the left and the hybrid nanowire part on the right. Two Pd (dark grey) gates are coupled to the nanowire via a dielectric $HfO_2$ (light grey) layer. The gate under the nanowire section with the superconducting shell (super gate) controls the electro-chemical potential in the hybrid. The gate under the bare nanowire section (tunnel gate) tunes a tunnel barrier at the semiconducting junction between the left Au lead and the hybrid. Four normal leads are tunnel-coupled to the hybrid and denoted as tunnel probes P0, P1, P2 and P3 in Fig. \ref{fig:1}B. The fifth lead forms a contact to the hybrid and is denoted as a drain lead. The tunnel probe P0 utilizes the semiconducting tunnel barrier controlled by the tunnel gate, while in the tunnel probes P1, P2 and P3 the Al oxide layer serves as a nm-thick tunnel barrier. The widths of probes P1, P2 and P3 are designed to be $200\,\mathrm{nm}$ and the lateral edge-to-edge distances between the neighboring probes are designed to be $200\,\mathrm{nm}$. Schematical transversal cross-sections of the device are displayed in Fig. \ref{fig:1}C-\ref{fig:1}E. The cross-section through the probes P1, P2 and P3 in Fig. \ref{fig:1}D indicates that the nanowire has the superconducting Al shell on one of its facets and that the Al oxide layer extends over the entire contact area between the hybrid and the Ag leads. The Al oxide layer in the drain area is removed by Ar ion milling prior to the deposition of the gold contacts - as shown in the cross-section through the drain lead in Fig. \ref{fig:1}E. Details of the steps in device fabrication can be found in the Device fabrication section and Fig. S1 in the Supplementary Materials.

A critical step in the fabrication of our hybrid nanowire devices is the formation of the Al oxide layer by an in-situ oxidation of the superconducting Al film. In order to test this fabrication step, we fabricate a planar tunnel junction which perpendicular cross-section is schematically shown in the top panel of Fig. \ref{fig:1}F. The junction leads are a superconducting Al (red) and a normal Ag (navy) film that partially overlap and that are separated by a thin dielectric Al oxide (pink) layer. The Al oxide is formed by an in-situ oxidation of the Al film, prior to the deposition of the Ag film. The tunnel junction is characterized in the bottom panel of Fig. \ref{fig:1}F by measuring the junction conductance as a function of a bias voltage $V_b$ and an in-plane magnetic field $B$. The result represents a typical tunneling spectroscopy of superconducting Al. As shown in the panel, coherence peaks spin-split with magnetic field due to the Zeeman effect. This demonstrates that the in-situ oxidation of Al can yield an Al oxide layer as a nm-thick tunnel barrier for tunneling spectroscopy. Next, we perform such in-situ oxidation on hybrid nanowires and characterize these hybrid nanowire devices in electrical transport measurements..

\subsection*{Characterization of tunnel probes and weak links}

We have studied three hybrid nanowire devices (Device 1, 2 and 3) that use nm-thick tunnel barriers. Devices 1 and 2 are nominally identical and are fully described in Fig. \ref{fig:1}. For Device 3, Al is deposited instead of Ag, so that three superconducting leads are defined on top of the Al oxide layer (see the Device fabrication section in the Supplementary Materials).  Therefore, the probes P1, P2 and P3 of Device 3 form three Josephson junctions with the hybrid nanowire. The replacement of Ag by Al in Device 3 is motivated by proposals that studying supercurrent in hybrid devices is an alternative way of detecting MZMs \cite{cxliu_prb_2021, schrade_arxiv_2018} and realizing MZM-based qubits \cite{schrade_prl_2018,schrade_prb_2017}.

\begin{figure}[!t] 
\centering
\includegraphics[width=\linewidth] {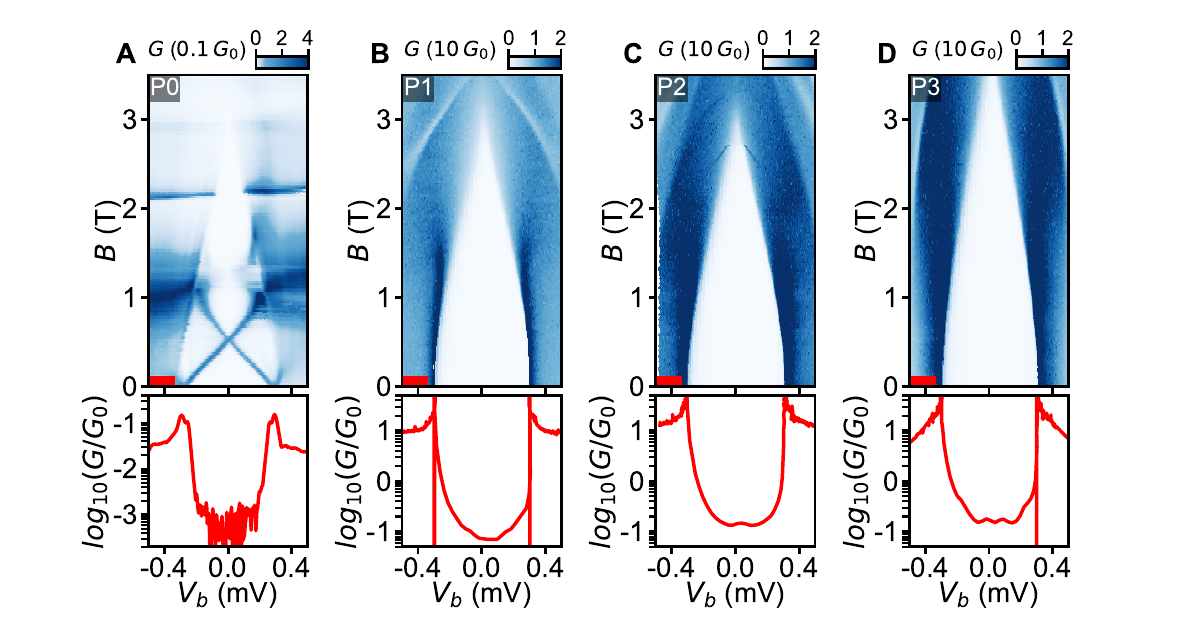}
\caption{\textbf{Tunnel probe characterization by tunneling conductance measurements}: Tunneling conductance $G$ as a function of a bias voltage $V_b$ and a parallel magnetic field $B$ along the nanowire of Device 1 measured by (\textbf{A}) probe P0, (\textbf{B}) probe P1, (\textbf{C}) probe P2 and (\textbf{D}) probe P3 in turn connected in Setup V1 (top row). The gate settings are $V_{TG}=2.1\,\mathrm{V}$ and $V_{SG}=0\,\mathrm{V}$. Red markers indicate the linecuts at zero field and corresponding traces are shown in the bottom row in logarithmic scale. By finding the coherence peak positions, a superconducting gap $\Delta$ is extracted to be (A) $290 \pm 6\,\mathrm{\mu eV}$, (B) $298 \pm 3\,\mathrm{\mu eV}$, (C) $305 \pm 11\,\mathrm{\mu eV}$ and (D) $301 \pm 5\,\mathrm{\mu eV}$.}\label{fig:2}
\end{figure}

As an initial step, the superconducting properties of Device 1 are characterized in conductance measurements by different probes in a voltage-bias setup. Each of the four probes P0, P1, P2 and P3 is in turn connected in Setup V1 to measure the tunneling conductance (see the Measurement setups section in the Supplementary Materials) and the results are shown in Fig. \ref{fig:2}. A voltage $V_{TG}=2.1\,\mathrm{V}$ is applied to the tunnel gate to define a tunnel barrier in the semiconducting junction of probe P0. The super gate is set to $V_{SG}=0\,\mathrm{V}$. In the top row of Fig. \ref{fig:2}, the tunneling conductance $G$ is measured by each probe as a function of a bias voltage $V_b$ and a parallel magnetic field $B$ along the nanowire of Device 1. In the parallel $B$-field, superconducting gaps detected by all four probes close at $3-3.5\,\mathrm{T}$. The variations in critical fields can be explained by small misalignments of the applied fields in different probes - as the nanowire is not perfectly straight. Furthermore, in comparison with the tunneling spectroscopy of the Al film in Fig. \ref{fig:1}F, generally there is no splitting of the coherence peaks measured by probes P1,  P2 and P3. This is most likely due to spin-mixing mechanisms induced by the spin-orbit interaction from the semicoducting nanowire \cite{bommer_prl_2019}. The phenomenon of non-splitting coherence peaks is widely seen in tunneling spectroscopies with semiconducting junctions - as shown here in Fig. \ref{fig:2}A. From each 2D-map in Fig. \ref{fig:2}, a linecut at zero field is taken and shown in logarithmic scale as a red trace in the bottom row of Fig. \ref{fig:2}. In some traces, large negative values appear at the coherence peaks and their origin is explained in the Measurement setups section in the Supplementary Materials. A superconducting gap $\Delta \sim 300\,\mathrm{\mu eV}$ of the hybrid can be obtained by extracting the positions of the coherence peaks. Noticeably, in all four probes the tunneling conductance at $V_b<\Delta$ (in-gap conductance) is roughly two orders of magnitude lower than the tunneling conductance at $V_b>\Delta$ (out-of-gap conductance). However, the out-of-gap conductance in these probes is two orders of magnitude larger than in probe P0 – due to the nm-thick tunnel barriers allowing for tunnel coupling of a large number of modes in the metallic leads of P1, P2 and P3. In order to measure the high out-of-gap conductance by P1, P2 and P3, the measurement sensitivity is adjusted, and consequently the modulations of the in-gap conductance cannot be precisely detected in Fig. \ref{fig:2}B-\ref{fig:2}D. The subgap spectra in the probes P1, P2 and P3 is in detail studied in Fig. \ref{fig:4} and Fig. \ref{fig:5} of this work. A characterization measurement like the one of Device 1 in Fig. \ref{fig:2} has been performed for Device 2 and similar results are shown in Fig. S4 in the Supplementary Materials.

\begin{figure}[!t] 
\centering
\includegraphics[width=0.5\linewidth] {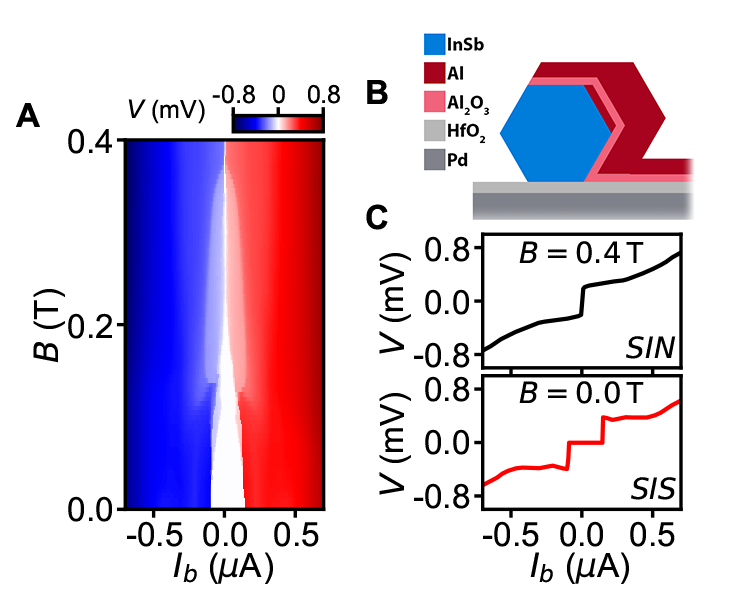}
\caption{\textbf{Tunnel probe characterization by supercurrent measurements}: Current-bias measurement of Device 3 with probes P1 and P2 connected in Setup I2 and probe P1 being current-biased. (\textbf{A}) Voltage drop $V$ as a function of an applied bias current $I_b$ and a parallel magnetic field $B$ along the nanowire. (\textbf{B}) Schematical transversal cross-section through the probes P1, P2 and P3 of Device 3 with superconducting Al (red) leads (the colors are as in Fig. \ref{fig:1}). (\textbf{C}) Linecuts from (A) taken at $B=0\,\mathrm{T}$ (bottom) and $B=0.4\,\mathrm{T}$ (top). The bottom linecut shows a switching current of $\sim 200\,\mathrm{nA}$ at zero field in the tunnel junction of probe P1 and corresponds to the SIS transport regime. The top linecut indicates the SIN transport regime - as the thick Al of the lead turns normal at high fields.}\label{fig:3}
\end{figure}

In order to test the Al oxide layer as a weak link for supercurrent measurements, current-bias measurements are performed on Device 3  and the results are shown in Fig. \ref{fig:3}. Fig. \ref{fig:3}B is a schematical cross-section through the probe P1 (or P2, or P3). It is shown that the probe P1, as well as P2 and P3, uses a superconducting lead made of thick Al. Together with the underlying Al oxide layer and the superconducting Al shell on the nanowire, the three superconducting leads of P1, P2 and P3 form three asymmetric Josephson junctions – JJ1, JJ2 and JJ3. In order to characterize JJ1, probes P1 and P2 are connected in Setup I2 and probe P1 is current-biased (see the Measurement setups section in the Supplementary Materials). A voltage drop $V$ over JJ1 is measured as a function of a bias current $I_b$ and a parallel magnetic field $B$ and the dependence is shown in Fig. \ref{fig:3}A. The linecuts taken at $B=0\,\mathrm{T}$ and $B=0.4\,\mathrm{T}$ are displayed in Fig. \ref{fig:3}C. The linecut taken at $B=0\,\mathrm{T}$ (bottom panel of Fig. \ref{fig:3}C) shows a zero voltage plateau due to the non-dissipative Josephson supercurrent with a switching current of $\sim 200\,\mathrm{nA}$. This demonstrates that at low fields the probe P1 is in the SIS transport regime (S-thin superconducting Al shell, I-thin dielectric Al oxide, S-thick superconducting Al lead). As the field increases in Fig. \ref{fig:3}A, the zero voltage plateau shrinks and disappears at $B \sim 0.2\,\mathrm{T}$ due to the suppressed superconductivity in the thick Al lead. Consequently, the SIS transport regime is altered by the SIN transport regime as the thick Al lead changes from being superconducting (S) to being normal (N). The linecut taken at $B=0.4\,\mathrm{T}$ (top panel of Fig. \ref{fig:3}C) confirms this as it resembles an $I-V$ characteristic of the tunneling transport between a superconductor and a normal metal. This suggests that a parallel field of $0.4\,\mathrm{T}$ is sufficient to turn the thick Al lead fully normal and that at high fields the probes P1, P2 and P3 of Device 3 can be used as normal metal leads for tunneling spectroscopy.

In Fig. \ref{fig:2} and Fig. \ref{fig:3} we demonstrate that the probes with nm-thick tunnel barrier can serve to characterize superconductivity in hybrid nanowires both in tunneling conductance measurements and supercurrent measurements. In the rest of this work, we focus on measuring in-gap conductance by different probes with a goal to study subgap states in hybrid nanowire devices.

\subsection*{Comparison between the tunneling spectroscopies by probes P0 and P1}

\begin{figure}[!t] 
\centering
\includegraphics[width=\linewidth] {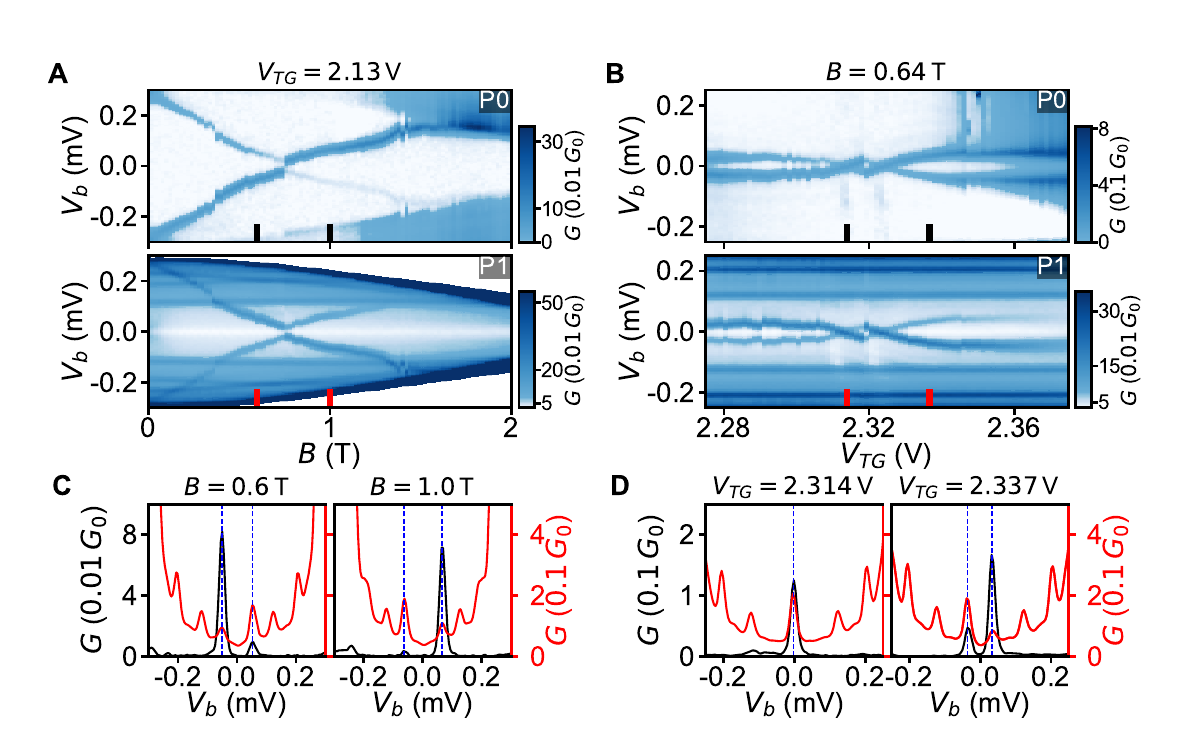}
\caption{\textbf{Tunneling spectrosopy by probes P0 and P1}: Conductance measurements of Device 1 with the probes P0 and P1 connected in Setup V2. (\textbf{A}) Tunneling conductance $G$ as a function of a bias voltage $V_b$ and a parallel magnetic field $B$. Tunnel gate is $V_{TG}=2.13\,\mathrm{V}$. (\textbf{B}) Tunneling conductance $G$ as a function of a bias voltage $V_b$ and a tunnel gate voltage $V_{TG}$. Parallel magnetic field is $B=0.64\,\mathrm{T}$. (\textbf{C}) Linecuts taken in (A) in black (probe P0) and red (probe P1) at the $B$ settings denoted by the markers and written as insets in (C). (\textbf{D}) Linecuts taken in (B) in black (probe P0) and red (probe P1) at the $V_{TG}$ settings denoted by the markers and written as insets in (D). Dashed vertical blue lines in (C) and (D) marked the conductance peaks corresponding to the same subgap state detected by both P0 and P1.}\label{fig:4}
\end{figure}

The capability of probes with nm-thick tunnel barriers to detect subgap states is examined for Device 1 in Fig. \ref{fig:4}. In-gap conductance is measured by two tunnel probes - probe P0 that utilizes the semiconducting tunnel barrier and probe P1 as the nearest probe that utilizes the nm-thick tunnel barrier. Probes P0 and P1 are connected in Setup V2 (see the Measurement setups section in the Supplementary Materials) and the super gate voltage is set at $V_{SG}=0\,\mathrm{V}$. In-gap conductance is measured by both probes as a function of a bias voltage $V_b$ and a parallel magnetic field $B$ (Fig. \ref{fig:4}A) or a tunnel gate voltage $V_{TG}$ (Fig. \ref{fig:4}B). Upon setting a value of $B$ or $V_{TG}$, $V_b$ is first swept on probe P0 with probe P1 at zero bias voltage and then $V_b$ is swept on probe P1 with probe P0 at zero bias voltage. The conductance dependences in the top panels of Fig. \ref{fig:4}A and \ref{fig:4}B show that a single subgap state is detected by probe P0 for the given ranges of $B$ and $V_{TG}$. The strong modulation by $V_{TG}$ (Fig. \ref{fig:4}B top) suggests that the subgap state is localized close to the semiconducting junction of probe P0. Such subgap states are commonly detected in tunneling spectroscopy measurements with semiconducting tunnel barriers in two-terminal \cite{demoor_njphys_2018} and three-terminal \cite{mazur_advmater_2022,vanloo_arxiv_2022,wang_prb_2022, anselmetti_prb_2019, menard_prl_2020} hybrid nanowire devices. Interestingly, the conductance dependences in the bottom panels of Fig. \ref{fig:4}A and \ref{fig:4}B show that the same subgap state is also detected by probe P1. This is additionally demonstrated by the linecuts taken from Fig. \ref{fig:4}A (Fig. \ref{fig:4}B) and displayed in Fig. \ref{fig:4}C (Fig. \ref{fig:4}D) in which aligned conductance peaks correspond to the same subgap state detected by the two probes. In addition, there are conductance peaks measured by probe P1 that are not measured by probe P0 – indicating that these subgap states most likely reside near P1 and are decoupled from P0. An additional tunnel gate sweep at a finite $B$-field and positive super gate shows that the subgap states detectable by both P0 and P1 remain detectable by P1 even when the semiconducting junction is pinched-off (see Fig. S5 in the Supplementary Materials). This means that the probe P1 can substitute the probe P0 when probing the hybrid nanowire end and even allow for tunneling spectrosopy in broader parameter ranges that are inaccessible to P0. An analogous measurement to the one in Fig. \ref{fig:4} has been carried out on Device 2 (see Fig. S6 in the Supplmentary Materials) and the capability of Al oxide tunnel probes to detect hybrid states is validated there as well. As we demonstrate that tunnel probes utilizing nm-thick tunnel barriers can detect subgap states in hybrid nanowires, in the rest of this work we use only these probes to study the subgap spectra in our hybrids.

\subsection*{Longitudinal dependence of subgap spectra studied by probes P1, P2 and P3}

\begin{figure} 
\centering
\includegraphics[width=\linewidth] {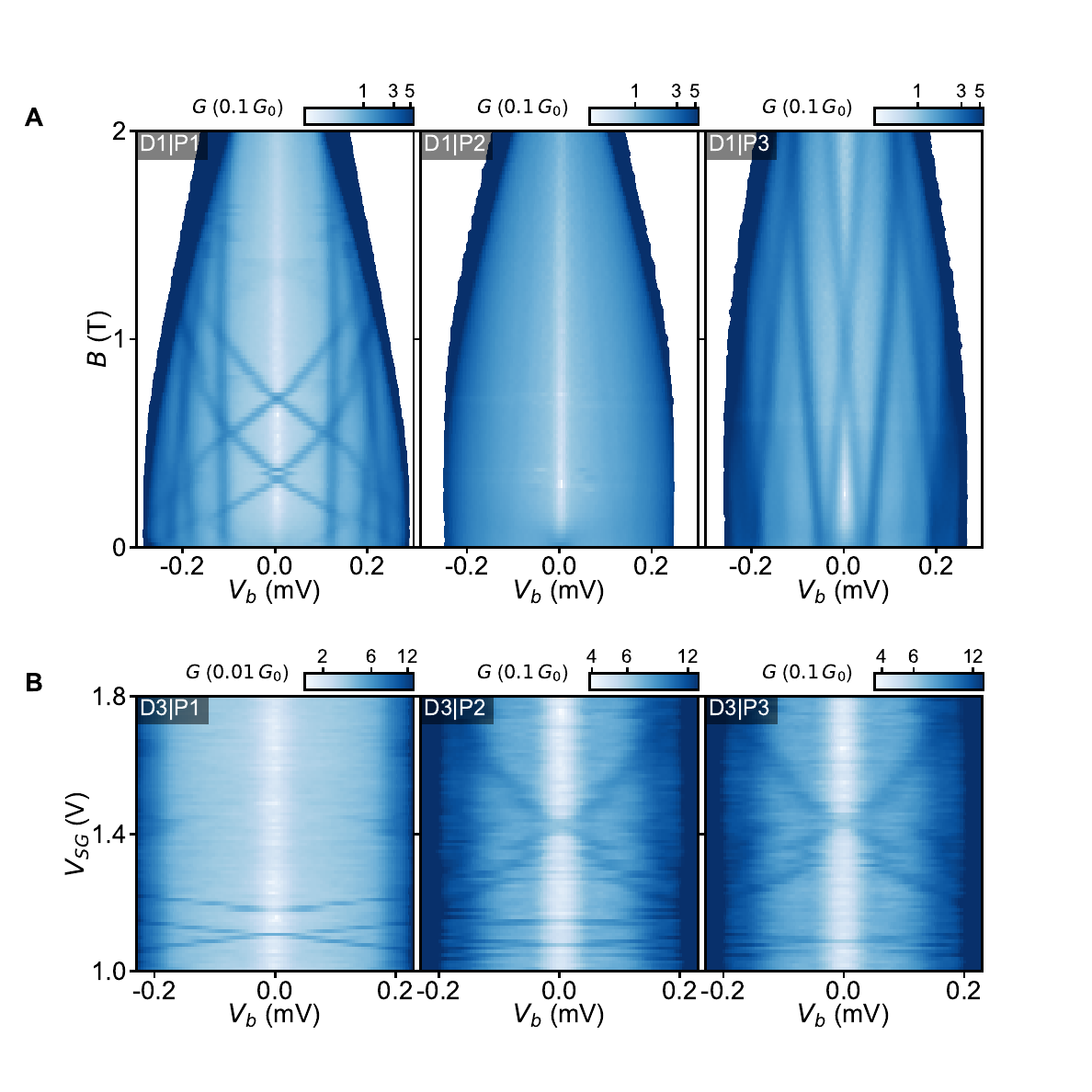}
\caption{\textbf{Tunneling spectrosopy by probes P1, P2 and P3}: (\textbf{A}) Tunneling conductance $G$ as a function of a bias voltage $V_b$ and a parallel magnetic field $B$ along the nanowire of Device 1. First, probes P1 (left) and P2 (middle) are connected in Setup V2 and then probes P1 and P3 (right) are connected in Setup V2. The super gate voltage is $V_{SG}=0.6\,\mathrm{V}$ and the tunnel gate is floated. All the subgap states are detectable only by single probes. (\textbf{B}) Tunneling conductance $G$ as a function of a bias voltage $V_b$ and a super gate voltage $V_{SG}$ of Device 3 measured by probes P1, P2 and P3 in-turn connected in Setup V1. A parallel magnetic field of $1\,\mathrm{T}$ is applied along the nanowire and the tunnel gate is floated. There is a  subgap state detectable by the probes P2 and P3, and non-detectable by the probe P1.}\label{fig:5}
\end{figure}

An appealing advantage of the tunnel probes with nm-thick Al oxide barriers is the opportunity to use multiple probes along a single hybrid nanowire for exploring the spatial distribution of subgap states. In Fig. \ref{fig:5} tunneling spectroscopy is performed by the probes P1, P2 and P3 of Device 1 and Device 3 in order to study the subgap spectra at different positions along the hybrid nanowires. The three probes of Device 1 are in pairs consecutively connected in Setup V2 (see the Measurement setups section in the Supplementary Materials) and the tunneling spectroscopy results are shown in Fig. \ref{fig:5}A. For Device 3, the three probes are in turn connected in Setup V1 (see the Measurement setups section in the Supplementary Materials) and the tunneling spectroscopy results are shown in Fig. \ref{fig:5}B. A high magnetic field is applied for the measurements of Device 3 in order to fully suppress the superconductivity in thick Al leads of the probes P1, P2 and P3.

The measurement in Fig. \ref{fig:5}A is performed with the super gate of Device 1 at $V_{SG}=0.6\,\mathrm{V}$ and the tunnel gate floated. Tunneling conductance is measured by probes P1, P2 and P3 as a function of a bias voltage $V_b$ and a parallel magnetic field $B$. The subgap spectra obtained by the three probes show different evolution with $B$ field. By probe P1, two kinds of subgap states are detected – subgap states insensitive to the $B$ field and subgap states with high $g$ factor that cross zero energy as the $B$ field is increased (Fig. \ref{fig:5}A left). The measurement in Fig. S5 in the Supplementary Materials demonstrates that the subgap states detected by the P1 reside at the hybrid end even when the semiconducting junction is pinched-off. This indicates that the states detected by the probe P1 are not junction states but subgap states localized at the end of the hybrid. 
These states appear to be strongly localized at the hybrid end, as no subgap states are detected by probe P2 (Fig. \ref{fig:5}A middle). Another subgap state with low $g$ factor is measured to be localized in the hybrid bulk - as it is detected by probe P3 (Fig. \ref{fig:5}A right), but is not detected by probe P2. The correlation between the states detected by different probes has been examined while varying the super gate of Device 1 (see Fig. S7 and Fig. S8 in the Supplementary Materials). The absence of correlations indicates that the subgap states detected in Device 1 are localized over less than $\sim 200\,\mathrm{nm}$ at the end or inside the bulk of the hybrid. A similar qualitative picture is observed for Device 2 and the corresponding measurements are shown in Fig. S9 in the Supplementary Materials. Besides confirming the strong localization of the subgap states in Device 1, the measurements of Fig. S7 and Fig. S8 show some additional features of the subgap states that can be used to better understand their nature. This is elaborated in the Disucssion section. 

For the measurements of Device 3, a parallel magnetic field $B=1\,\mathrm{T}$ is applied along the nanowire and the tunnel gate is floated. The superconductivity in the thick Al leads of probes P1, P2 and P3 is fully suppressed due to the high field, and they are used as tunnel probes. In Fig. \ref{fig:5}B, each of the three probes is in turn connected in Setup V1 (see the Measurement setups section in the Supplementary Materials) and the tunneling conductance is measured as a function of a bias voltage $V_b$ and a super gate voltage $V_{SG}$. The order in which the 2D maps are measured is P2, P1, P3 (middle, left, right in Fig. \ref{fig:5}B). Even the three 2D maps are obtained in three consecutive $V_{SG}$ sweeps, a striking similarity between the two subgap features detected by probes P2 and P3 indicate the presence of a subgap state coupled to both probes. However, the absence of any similar feature in the tunneling spectroscopy by probe P1 (taken in between the measurements by P2 and P3) suggests that the same state is not detectable at the hybrid nanowire end. This implies that the subgap state extends over more than $400\,\mathrm{nm}$ in the hybrid bulk, but does not reach the hybrid end. Importantly, detecting such state shows the capability of probes with nm-thick tunnel barriers to detect extended subgap states.  Another extended subgap state is detected in the same device in other $V_{SG}$ range (see Fig. S10 in the Supplementary Materials). Additional tunneling spectroscopy in a broad super gate range ($-10\,\mathrm{V}<V_{SG}<10\,\mathrm{V}$) in all three probes shows that applying positive $V_{SG}$ weakly reduces the superconducting gap (see Fig. S11A in the Supplementary Materials). At $V_{SG}=10\,\mathrm{V}$ (and $B=1\,\mathrm{T}$) the gap remains open along the hybrid of Device 3. Additional supercurrent measurement at zero field shows that sweeping $V_{SG}$ from $-2\,\mathrm{V}$ to $2\,\mathrm{V}$  has no effect on the supercurrent measured by probe P1 (see Fig. S11B in the Supplementary Materials). Together with the large switching current value, such insensitivity to the super gate indicates that the hybrid states have a negligible contribution to the supercurrent that is dominantly carried by the condensate in the Al shell.

\section*{Discussion}

In order to investigate the origin of various subgap states in our devices, their sensitivity to magnetic and electric fields is examined in several additional measurements shown in Figs. S7, S8 and S9 in the Supplementary Materials. We find that subgap states with high $g$ factor are sensitive to local electric fields (Figs. S7 and S9), while the subgap states with low $g$ factor are weakly sensitive or insensitive to local electric fields (Figs. S7 and S8). This is in agreement with the nature of hybrid states, where the $g$ factor of a state and its sensitivity to electric fields are both determined by its wavefunction distribution between the superconductor and semiconductor.

Subgap states with high $g$ factor are formed for sufficiently positive super gate (Fig. S7). Such states were previously interpreted as the bulk states inside an electric field-tunable hybrid. However, our measurements show that these states are not bulk states, as they are localized at the hybrid end. This shows that the edge of the superconducting shell significantly influences the effect of the super gate on the subgap spectrum. Namely, our result suggests that the proximity effect is weaker close to the edges of the superconducting shell and that localized subgap states with high $g$ factor may be inevitably present at the ends of hybrid nanowires. We also find that single detached subgap states can be present inside the hybrid bulk (Fig. S9). 

% Moreover, this suggests that localized subgap states with high $g$ factor may be inevitably present at the hybrid ends for positive super gate. We also find that single detached subgap states can be present inside the hybrid bulk (Fig.\ref{fig:d2p1p2tg}). 

% Particularly in Fig. \ref{fig:d1p1p2sg}, we observe that positive super gate induces subgap states with high $g$ factor coupled to the hybrid end. Such states were previously interpreted as electric field-tunable hybrid bulk states However, we find that these states are localized at the hybrid end 

% This is in agreement with the nature of hybrid states, where the $g$ factor of a state and its sensitivity to electric fields are both determined by its wavefunction distribution between the superconductor and semiconductor.

% The states with low $g$ are strongly coupled to the Al and therefore likely to be localized at the InSb-Al interface. There, the electric field from the gates is additionally strongly screened by the Al shell. Moreover, these states may be affected by a large spin-orbit field at the InSb-Al interface - as a large electric field at the interface can be caused by the band bending. Therefore, such large spin-orbit field could explain the insensitivity to $B$-field of some subgap states in Fig. \ref{fig:5}. The same mechanism has been used to explain the magnetic field-independent subgap states detected in InAs-Al hybrid nanowires \cite{junger_prl2020}. 

By the probes with nm-thick tunnel barriers, subgap states with low $g$ factors are observed and these states also show insensitivity to the gates. We speculate that these states may be formed at the InSb-Al interface. The electric field at the interface is strongly screened by the Al shell. Besides, strong spin-orbit interaction could be present at the interface due to the band bending - leading to the magnetic field-insensitivity of the interface states \cite{junger_prl2020}.

Most of the subgap states in our devices can be detected by only one tunnel probe ($\sim 200\,\mathrm{nm}$ extension) - either at the hybrid end (by probe P1) or inside the hybrid bulk (by probe P2 or P3) - while some subgap states can be detected by two tunnel probes ($\sim 400\,\mathrm{nm}$ extension). However, we do not report any subgap state being detectable by all three tunnel probes ($\sim 600\,\mathrm{nm}$ extension). This is comparable with the results of a previous study \cite{grivnin_natcomm_2019}, where tunnel probes had lateral separations of $\sim 500\,\mathrm{nm}$ and there was no report on subgap states detected by multiple probes. The presence of the localized subgap states and the absence of extended bulk subgap states can be caused by inhomogeneities in the electro-chemical potential due to disorder in the hybrid nanowires \cite{pan_prr_2020}. Potentially, additional disorder in our devices can originate from the formation of the tunnel probes as their leads may induce additional stress on the nanowires. However, we emphasize that the tunneling spectroscopies done by the gate-defined tunnel probes in our devices are fully comparable to the analogous measurements on standard two-terminal and three-terminal InSb-Al hybrid nanowire devices without nm-thick Al oxide probes.

A recent work on three-terminal nanowire hybrids has used non-local measurements to study the hybrid bulk \cite{vanloo_arxiv_2022}. There, finite non-local conductance signals arising at low bias voltages and high positive super gate voltages have been interpreted as closing of an induced superconducting gap in the hybrid bulk due to an electrostatic effect of the super gate. In our work, however, no gap-closing at positive super gate voltages is detected in the hybrid bulk. A possible reason for this is that the bulk states giving rise to the non-local signals are nanowire states that are weakly coupled or even non-coupled to the superconductor. Therefore, such dominantly semiconducting states would weakly contribute or even may not contribute to the tunneling spectroscopy signals in our work which are detected through the superconducting shell. 

\section*{Conclusion}

In conclusion, we develop a new type of tunnel probes for tunneling spectroscopy of hybrid InSb-Al nanowires. These probes use a nm-thick layer of Al oxide as a tunnel barrier that is created by in-situ oxidation of the superconducting Al shell on the nanowires. Normal or superconducting leads defined by shadow-wall lithography technique on top of the Al oxide layer are used to probe the nanowire hybrids in tunneling conductance and supercurrent measurements. We demonstrate that such probes provide an alternative way of measuring subgap spectra at the nanowire ends, and therefore can replace standardly used tunnel probes defined by local gates. Particularly, the nm-thick tunnel barrier can significantly diminish unwanted smooth potential effects, which inevitably exist in devices with semiconducting junctions. Furthermore, the tunnel probes with Al oxide tunnel barriers can be defined at any position along a hybrid nanowire and therefore can be used to directly probe the hybrid bulk. We exploit this advantage and utilize these tunnel probes to study the longitudinal dependence of the subgap spectra in multiple hybrid nanowires. As a result, we identify Andreev bound states of various extension at the ends and inside the bulks of the hybrids. Our work offers a new way of investigating the bulk-edge correspondence in superconducting-semiconducting nanowires.    

\section*{Data and code availability}

Raw data and codes for the data analysis and plotting of the figures can be found at:\\ \url{https://doi.org/10.5281/zenodo.7662232}.

\section*{Conflicts of interest}

The authors declare no conflicts of interest.

\section*{Author contributions}

J.-Y. W., V. L. and L. P. K. conceived the experiment. G. B., S. G. and E. P. A. M. B. conducted the nanowire growth. J.-Y. W., G. P. M., C. S., M. L., J. C. W. and A. B. contributed to different steps in the device fabrication. V. L. and J.-Y. W. carried out the transport measurements and performed the data analysis. V. L., J.-Y. W. and L. P. K. prepared the manuscript with comments from the other co-authors. J.-Y. W. and L. P. K. supervised the project.   

\section*{Acknowledgments}

The authors thank Anton R. Akhmerov and Tom Dvir for valuable discussions, and thank Nick van Loo for valuable comments on the manuscript. The authors are also greatful to Olaf Benningshof and Jason Mensingh for technical support. This work was financially supported by the Dutch Organization for Scientific Research (NWO) and the Foundation for Fundamental Research on Matter (FOM).   

\bibliography{scibib}
\bibliographystyle{Science}

\clearpage

\end{document}

\baselineskip24pt
\maketitle 

\section*{Device fabrication}

\begin{figure}[!t] 
\centering
\includegraphics[width=\linewidth] {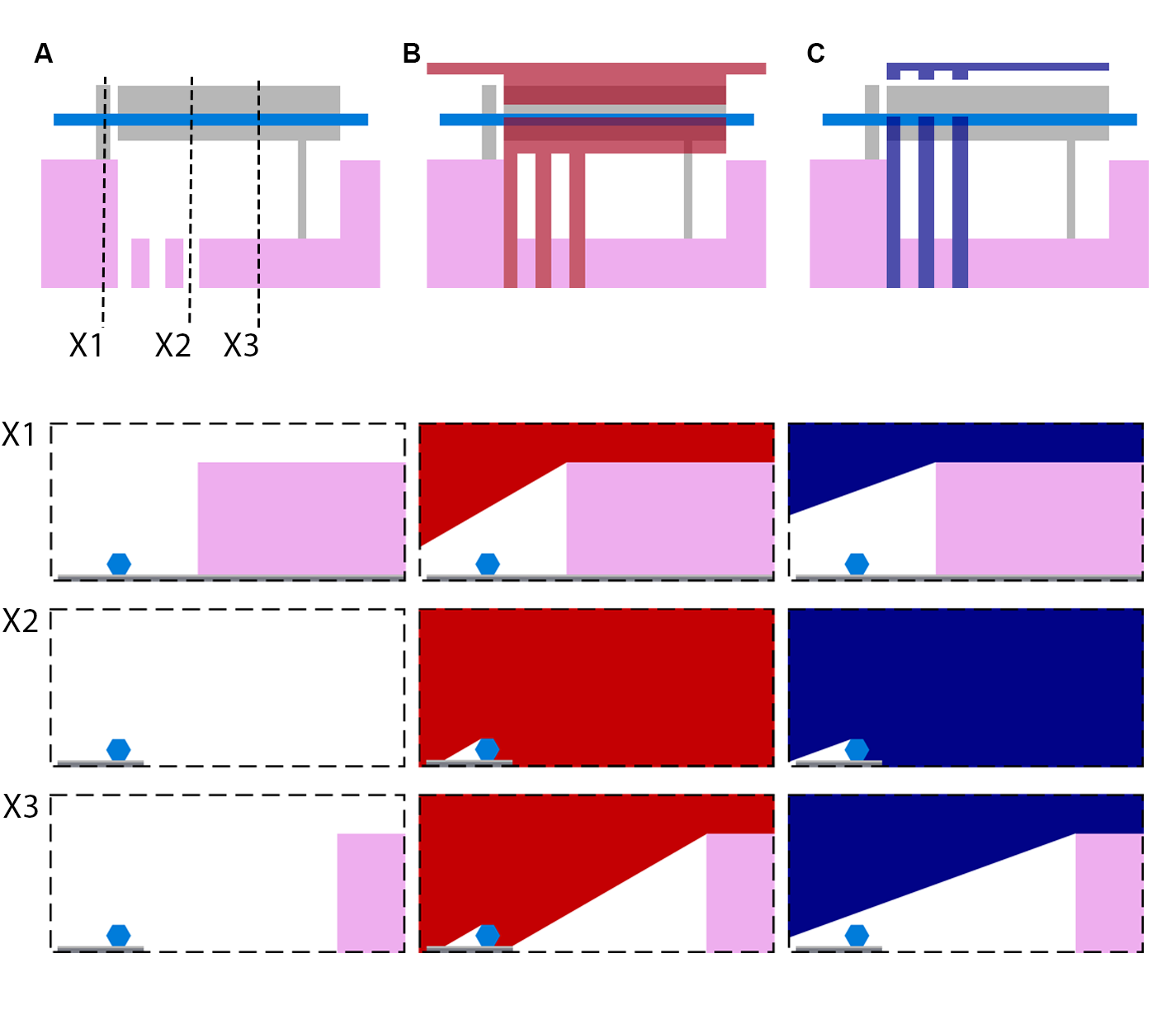}
\caption{\textbf{Device fabrication by shadow-wall lithography}: (\textbf{A}) A schematical representation of the top view on a chip with nanowire (light blue), gates (grey) and shadow-walls (lilac). Dashed black lines denote three perpendicular cross sections X1, X2 and X3 depicted below. (\textbf{B}) A schematical representation of the shadowing effects of the shadow-walls in the growth of Al (red) at $30^{\,\circ}$ with the corresponding schematics through the cross-sectional cuts X1, X2 and X3 below. (\textbf{C}) A schematical representation of the shadowing effects of the shadow-walls in the growth of Ag (navy) (Device 1 and 2) or thick Al (Device 3) at $18^{\,\circ}$ with the corresponding schematics through the cross-sectional cuts X1, X2 and X3 below.}\label{fig:fab}
\end{figure}

For the devices in this work, intrinsic Si wafers covered with $285\,\mathrm{nm}$ of thermal $\mathrm{SiO_2}$ were used as substrates. On top of the $\mathrm{SiO_2}$ layer, local gates (tunnel gate and super gate) were lithographically defined and grown by electron-beam evaporation by depositing $3/17\,\mathrm{nm}$ of Ti/Pd. This step was followed by an atomic layer deposition (ALD) step in which $\sim20\,\mathrm{nm}$ layer of high quality $\mathrm{HfO_2}$ was grown at $110^{\circ}$C to serve as a dielectric of the gates. The shadow-wall structure was lithographically defined on top of the ALD $\mathrm{HfO_2}$. First, FOx-25 (HSQ) was spun at $1.5\,\mathrm{krpm}$ for 1 minute and hot-baked at $180^{\circ}$C for 2 minutes. The HSQ layer was then lithographically patterned and developed with MF-321 at $60^{\circ}$C for 5 minutes. After the formation of the HSQ shadow-walls, stemless InSb nanowires were precisely deposited by an optical nanomanipulator setup on top of the gates.

The deposition of the superconducting Al film and its controlled in-situ oxidation were performed in the same way for all nanowire devices in this work. First, the native oxide was removed from the nanowire surface by gentle hydrogen cleaning. The superconducting Al film was then grown on top of the InSb nanowire surface by directional evaporation of Al at a temperature of $140\,\mathrm{K}$. The Al was deposited at an angle of $30^{\circ}$ with respect to the substrate plane and with a thickness of $5.5\,\mathrm{nm}$. Due to the regular hexagonal nanowire cross-section and the specific deposition angle, the Al film is deposited on three nanowire facets. On one facet, the Al film is perpendicularly grown and therefore has $5.5\,\mathrm{nm}$ thickness after the deposition. The direction of the Al forms angles of $30^{\circ}$ with the other two facets and the substrate, so that the Al film thickness there after the deposition is a half of the flux - $\sim2.5\,\mathrm{nm}$. The Al growth was followed by the in-situ oxidation in the load lock of the evaporator. Here, the Al film was oxidized for 10 minutes at $10\,\mathrm{Torr}$ oxygen pressure. This was precisely controlled such that the Al film on one nanowire facet (where it is thicker) is partially oxidized and the Al films on the other two nanowire facets (where it is thinner) and on the substrate are fully oxidized. As a result, the superconducting Al film of $\sim3\,\mathrm{nm}$ thickness remains only on one nanowire facet and is covered by $\sim2,5\,\mathrm{nm}$ of dielectric $\mathrm{Al_2O_3}$ layer. The layer of $\mathrm{Al_2O_3}$ continuously extends to the other two neighboring facets and the substrate, as the Al there is completely turned into $\mathrm{Al_2O_3}$. The full oxidation of Al on the substrate was confirmed by measuring high ($\sim \mathrm{G\Omega}$) resistance of the Al film after the oxidation on a test chip without nanowires. After the oxidation in the load lock, the sample was inserted back into the evaporation chamber, and it was warmed up to the room temperature. At the room temperature, the deposition of Ag (for Device 1 and 2) or Al (for Device 3) was executed with a thickness of $80\,\mathrm{nm}$ and at an angle of $18^{\circ}$ with respect to the substrate. Due to the smaller deposition angle in comparison to the initial Al deposition at $30^{\circ}$, the HSQ shadow-walls cast longer shadows and block the growth of Ag or thick Al on the nanowire. Consequently, the interruptions in the HSQ shadow-wall determine where the growth of the three Ag or three thick Al leads takes place along the hybrid on top of the $\mathrm{Al_2O_3}$ layer. These leads form continuous connections to the substrate and extend to pre-patterned bonding pads and are used as probes P1, P2 and. P3. 
In Fig. \ref{fig:fab} the effect of the HSQ shadow-walls in different depositions is schematically shown in order to clarify how the specific device layout is obtained.   

The normal probe P0 and the drain lead were fabricated ex-situ after the growth of Ag or thick Al leads, in the same way for all devices. These contacts were made by electron-beam evaporation of $10/120\,\mathrm{nm}$ of Ti/Au at two nanowire ends after Ar ion milling was used to remove the native oxide (for P0 lead) and $\mathrm{Al_2O_3}$ layer (for the drain lead).

For the fabrication of the planar tunnel junction introduced in Fig. 1F in the main text, the Al and Ag films were grown at an angle of $50^{\circ}$ and the in-situ oxidation pressure was $1\,\mathrm{Torr}$.

\section*{Measurement setups}

\begin{figure}[!t] 
\centering
\includegraphics[width=\linewidth] {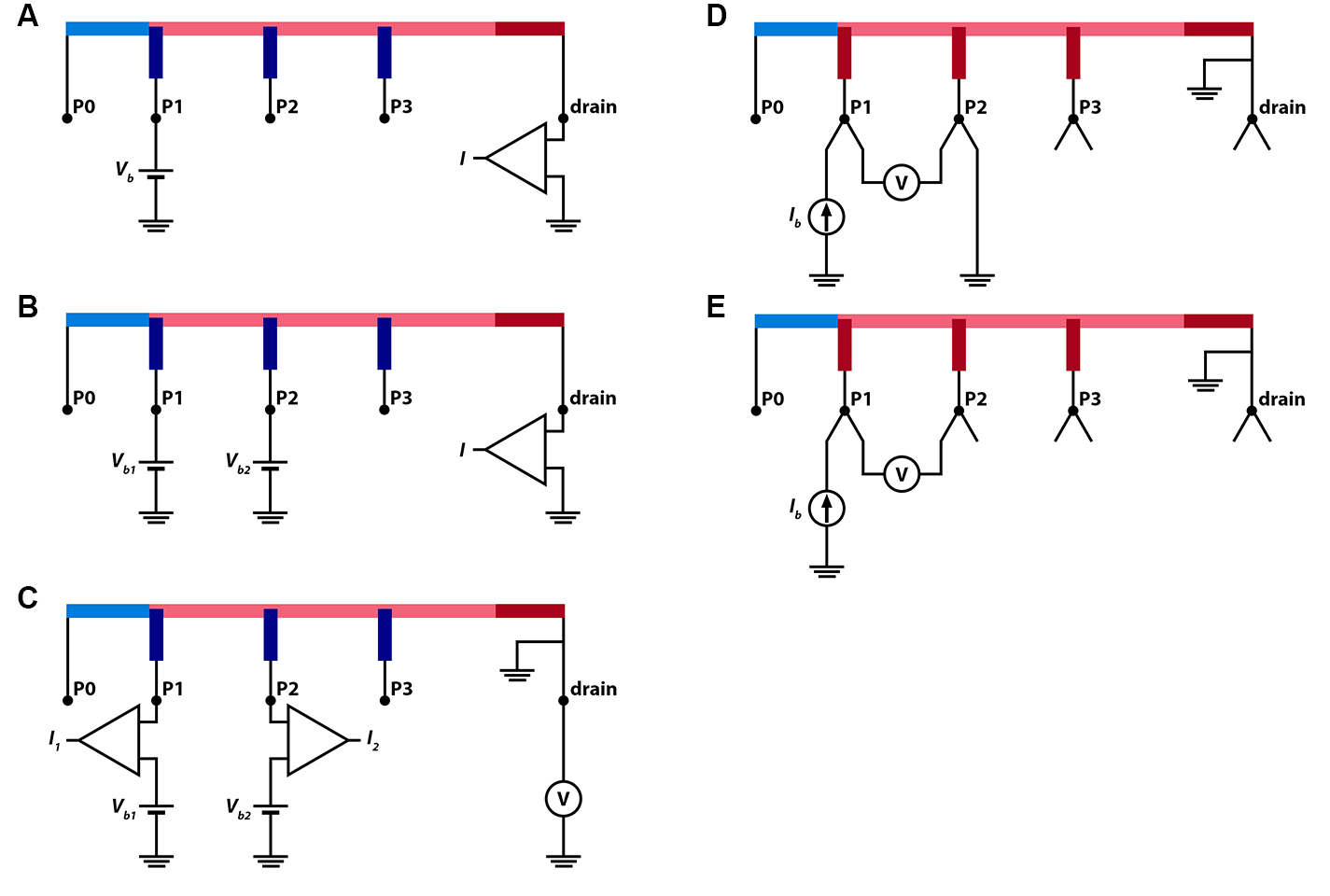}
\caption{\textbf{Measurement setups}: Each panel contains a schematical representation of the nanowire device with the four probes P0, P1, P2 and P3 and the drain lead (see Fig. 1 in the main text). The five panels represent: (\textbf{A}) probe P1 connected in Setup V1, (\textbf{B}) probes P1 and P2 connected in Setup V2, (\textbf{C}) probes P1 and P2 connected in Setup V3, (\textbf{D}) probes P1 and P2 connected in Setup I1 and (\textbf{E}) probes P1 and P2 connected in Setup I2 with probe P1 being current-biased.}\label{fig:setups}
\end{figure}

All measurements were performed at a base temperature of $\sim 20\,\mathrm{mK}$ inside a dilution refrigerator equipped with a superconducting vector magnet. The devices are designed and fabricated to be parallel to the magnet axis with the largest field limit and high parallel fields are therefore applied along this axis. However, small misalignments of the nanowires can be caused by possible small bending along a single nanowire or small variations in the nanowire orientation during its deposition on the substrate. Consequently, the parallel magnetic fields may be misaligned by few degrees from the nanowires. 

For the conductance measurements, three different voltage bias setups (Setup V1, V2 and V3) are used in the standard lock-in configuration. For the supercurrent measurements, two different current bias setups (Setup I1 and I2) are used in the four-terminal configurations. Table \ref{table:S1} shows an overview of the figures in this work and the setups used to obtain the corresponding data in the figures.  All setups are schematically shown in Fig . \ref{fig:setups} and in detail described in the following paragraphs.

\begin{table}
\begin{center}
\begin{tabular}{ |c|c| }
 \hline
 Figure & Setup \\
 \hline
 1 & V1 \\
 2 & V1 \\ 
 3 & I2 \\
 4 & V2 \\
 5a & V2 \\
 5b & V1 \\
 S3 & I1 \\
 S4 & V3 \\
 S5 & V3 \\
 S6 & V2 \\
 S7 & V2 \\
 S8 & V2 \\
 S9 & V3 \\
 S10 & V1 \\
 S11a & V1 \\
 S11b & I2 \\
 \hline
\end{tabular}
\caption{\label{table:S1}An overview of figures and corresponding measurement setups.}
\end{center}
\end{table}

In the voltage bias setups, dc-voltage sources are used to set dc-bias voltages and current-meters are used to measure dc-currents. Lock-in amplifiers are used to apply ac-voltages ($10\,\mathrm{\mu V}$ amplitude) and measure ac-currents in order to obtain differential conductance. The set dc-bias voltages and measured differential conductance values are corrected for voltage drops over a series resistance depending on the corresponding setup circuit.

Setup V1 represents a two-terminal voltage-bias setup where a bias voltage $V_b$ is applied to one probe and current $I$ is measured in the drain lead. All the remaining probes are floated. In Fig. \ref{fig:setups}A probe P1 is connected in Setup V1. The series resistance $R_s=8.89\,\mathrm{k\Omega}$ in this setup includes two fridge lines, series resistances of the voltage source and the current meter and low-pass filters on the PCB. When the conductance is high, even a slight overestimate of $R_s$ may lead to falsely measured negative differential conductance - as the ac-voltage drop due to the correction becomes $dV<0$. This is the reason for large negative values in the traces in Fig. 2 - since the negative $dI/dV$ values correspond to negative infinity in logarithmic scale.  

In Setup V2 bias voltages $V_{b1}$ and $V_{b2}$ are applied to two probes and current $I$ is measured in the drain lead. The remaining two probes are floated. Two lock-in amplifiers are used to set ac-components of $V_{b1}$ (lock-in1) and $V_{b2}$ (lock-in2). In Fig. \ref{fig:setups}B probes P1 and P2 are connected in Setup V2. Upon setting a parameter value (magnetic field or gate voltage), the bias voltages are consecutively swept on the two probes and differential conductance is measured by the corresponding lock-in amplifier. First, $V_{b1}$ is swept and the lock-in1 is used to measure the ac-current $di$ in the drain, while the other probe is kept at $V_{b2}=0\,\mathrm{V}$. Then, $V_{b2}$ is swept and the lock-in2 is used to measure the ac-current $dI$ in the drain, while the previous probe is inactive and kept at $V_{b1}=0\,\mathrm{V}$. In addition to setting the dc-voltage bias to zero when the probes are inactive, the corresponding lock-ins are also set at zero-frequency and minimal amplitude ($4\,\mathrm{nV}$), such that the inactive probe is effectively grounded. This results in a voltage divider effect such that the measured dc-current $I$ and ac-current $dI$ are underestimated. However, the resistance of the superconducting shell is $\sim 100$ times smaller than the resistance to the inactive probe – meaning that the voltage divider effect is negligible, and the currents are dominantly drained by the drain lead. This is also confirmed in Fig. \ref{fig:d1pinchoff} - where the conductance measured by one probe (P1) does not change upon changing the inactive probe (P0) from pinch-off to tunneling regime. With neglecting the voltage divider effect, the series resistance for each probe connected in Setup V2 is the same as in Setup V1 – $R_{s1}=R_{s2}=8.89\,\mathrm{k\Omega}$.

In Setup V3 bias voltages $V_{b1}$ and $V_{b2}$ are applied to two probes and currents $I_1$ and $I_2$ are measured in these probes while the drain lead is connected to the cold ground. Bias voltages are corrected for a cold ground potential measured by an additional voltmeter. The remaining two probes are floated. Two lock-in amplifiers are used to set ac-components of $V_{b1}$ (lock-in1) and $V_{b2}$ (lock-in2). In. Fig. \ref{fig:setups}C probes P1 and P2 are connected in Setup V3. As in Setup V2, upon setting a parameter value (magnetic field or gate voltage), $V_{b1}$ and $V_{b2}$ are consecutively swept on the two probes. Here, the series resistance for each probe is $R_{s1}=R_{s2}=5.81\,\mathrm{k\Omega}$ and includes series resistances of the voltage source and the current-meter and only one fridge line and one low-pass filter – as the drain lead is cold-grounded.

In the current bias setups, dc-current sources are used to set dc-bias currents and voltmeters are used to measure dc-voltages. In order to allow for four-terminal. Configurations, each lead is bonded to two fridge lines. The drain lead is also bonded to the cold ground. In the current-bias measurements probe P0 is always floated.

In Setup I1 bias current $I_b$ is applied between two leads (probe lead or drain lead) and the voltage drop $V$ measured between these two probes. The remaining leads are floated. In Fig. \ref{fig:setups}D probes P1 and P2 are connected in Setup I1. Note that this setup measured a series of two Josephson junctions if a pair of probes P1, P2 and P3 is connected. 

In Setup I2 bias current $I_b$ is applied from one probe to the drain and the voltage drop $V$ is measured between that probe and its first neighboring probe. The remaining probes are floated. In Fig. \ref{fig:setups}E probes P1 and P2 are connected in Setup I2 and the probe P1 is current-biased. Note that in this configuration the voltage drop on a potential residual contact resistance in the drain contact is not measured. 

\begin{figure}[!t] 
\centering
\includegraphics[width=\linewidth] {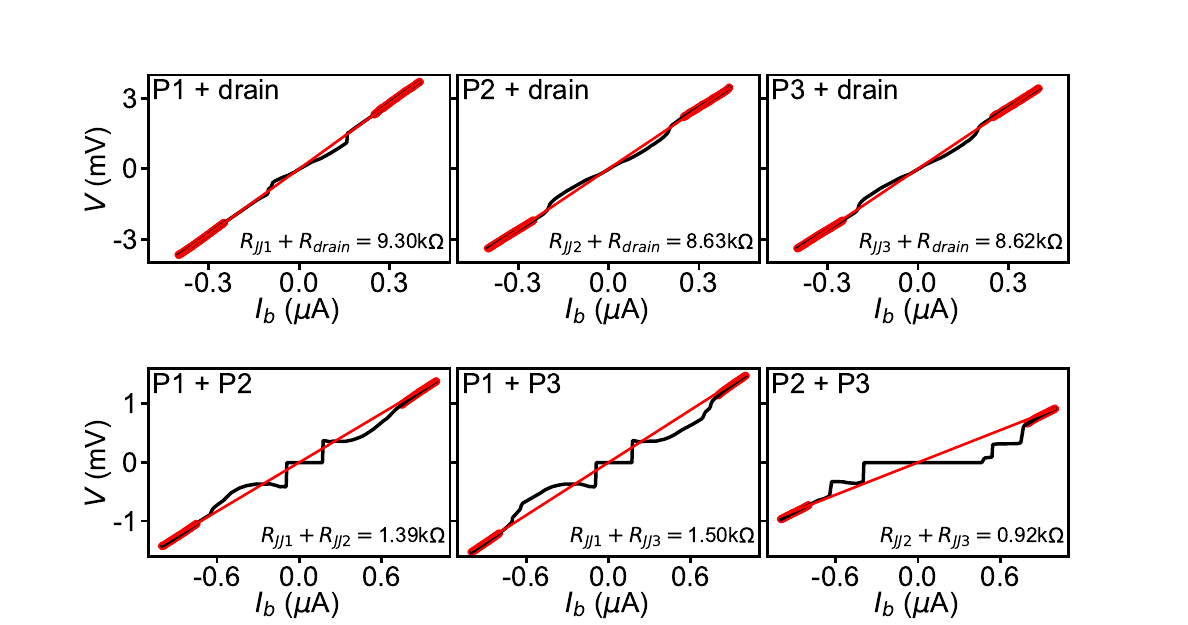}
\caption{\textbf{Contact resistance of the drain in Device 3}: $I-V$ characteristics (black) obtained in current-bias measurements of Device 3 at zero field. Corresponding pairs of leads connected in Setup I1 are denoted in the upper-left corners of the six panels. By fitting the linear parts of the characteristics at high bias (red), a series of  two Josephson junctions or of one Josephson junction and drain  is fitted and denoted in  the bottom-right corners of six panels. The resistances of the three Josephson junctions and the resistance in the drain contact are estimated to be: $R_{JJ1} \sim 1.1\,\mathrm{k\Omega}$, $R_{JJ2} \sim 0.4\,\mathrm{k\Omega}$, $R_{JJ3} \sim 0.5\,\mathrm{k\Omega}$ and $R_{drain} \sim 8.2\,\mathrm{k\Omega}$,.}\label{fig:rdrain}
\end{figure}

Device 3 is first characterized at zero field by measuring $I-V$ characteristics when each pair of its leads is connected in Setup I1. If the two leads are probe leads, a series of two Josephson junctions is measured (in Fig. \ref{fig:setups}d this would be a series of JJ1 and JJ2). If one lead is a probe lead and the other lead is the drain lead, a single Josephson junction is ideally measured. However, these measurements revealed a residual series resistance of $R_{drain}\sim8.2\,\mathrm{k\Omega}$ in the drain contact, as shown in Fig. \ref{fig:rdrain}. Therefore, the current-bias measurements of Device 3 in the main text are performed in Setup I2 that excludes this resistance. This resistance can be attributed to an incomplete removal of $\mathrm{Al_2O_3}$ by the Ar ion milling in the ex-situ fabrication of the drain lead. In order to avoid voltage divider effects due to the resiaul resistance, the. conductance measurements of Device 3 in the main text are performed in Setup V1 rather than in Setup V2 or V3. In Setup V1 the residual contact resistance is much smaller than the subgap tunneling resistance - meaning that the bias voltage dominantly drops on the tunnel junctions of the ptobes P1, P2 and P3.

\section*{Extended data}

\begin{figure}[!t] 
\centering
\includegraphics[width=\linewidth] {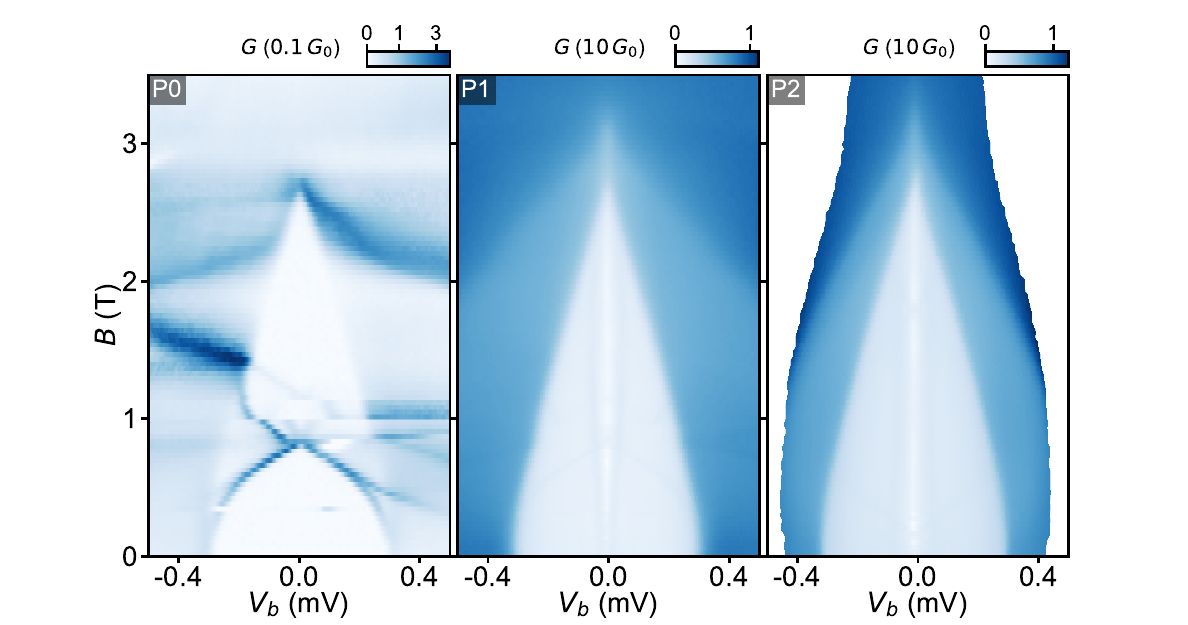}
\caption{\textbf{Tunnel barrier characterization by tunneling conductance measurements (Device 2)}:  Tunneling conductance $G$ as a function of a bias voltage $V_b$ and a parallel magnetic field $B$ measured by the probes P0, P1 and P2. First, probes P0 (left) and P1 (middle) are connected in Setup V3 and then probes P1 and P2 (right) are connected in Setup V3. The probe P3 of Device 2 is not functional. Voltages at the tunnel gate and super gate are $V_{TG}=0.75\,\mathrm{V}$ and $V_{SG}=0\,\mathrm{V}$, respectively. }\label{fig:d2gap}
\end{figure}

\begin{figure}[!t] 
\centering
\includegraphics[width=0.5\linewidth] {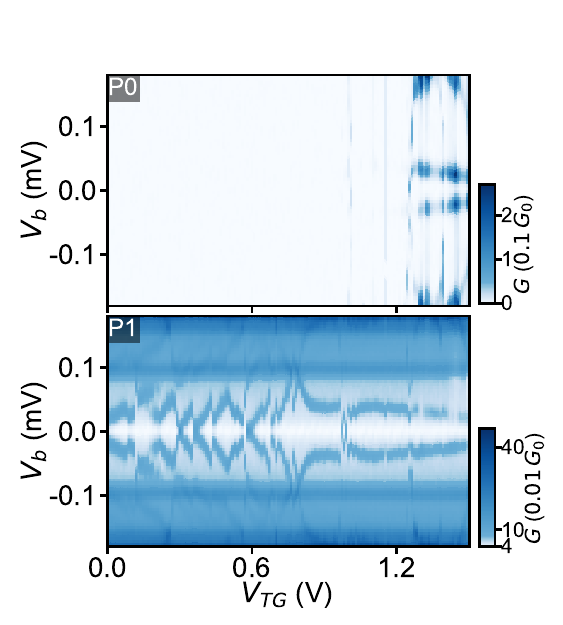}
\caption{\textbf{Effect of the tunnel gate on the tunneling spectroscopy by probes P0 and P1 (Device 1)}: Tunneling conductance $G$ as a function of a bias voltage $V_b$ and a tunnel gate voltage $V_{TG}$ measured by the probes P0 and P1 connected in Setup V2. The super gate voltage is $V_{SG}=0.6\,\mathrm{V}$ and a magnetic field of $0.23\,\mathrm{T}$ is applied perpendicular to the substrate plane.}\label{fig:d1pinchoff}
\end{figure}

\begin{figure}[!t] 
\centering
\includegraphics[width=0.5\linewidth] {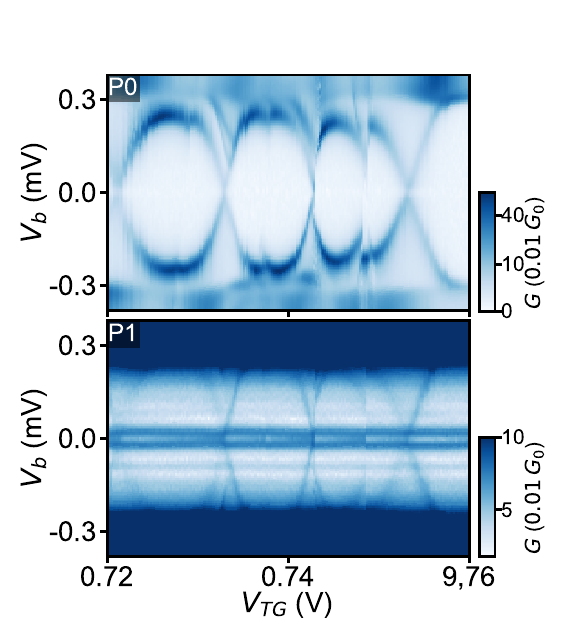}
\caption{\textbf{Tunneling spectrscopy by probes P0 and P1 (Device 2)}: Tunneling conductance $G$ as a function of a bias voltage $V_b$ and a tunnel gate voltage $V_{TG}$ measured by probe P0 (top) and P1 (bottom) connected in Setup V3. Magnetic field is $B=0\,\mathrm{T}$ and the super gate voltage is $V_{SG}=0\,\mathrm{V}$. Subgap states tunable by the tunnel gate are detectable by both probes P0 and P1. Subgap states insensitive to the tunnel gate are only detectable by probe P1.}\label{fig:d2p0p1}
\end{figure}

\begin{figure}[!t] 
\centering
\includegraphics[width=0.5\linewidth] {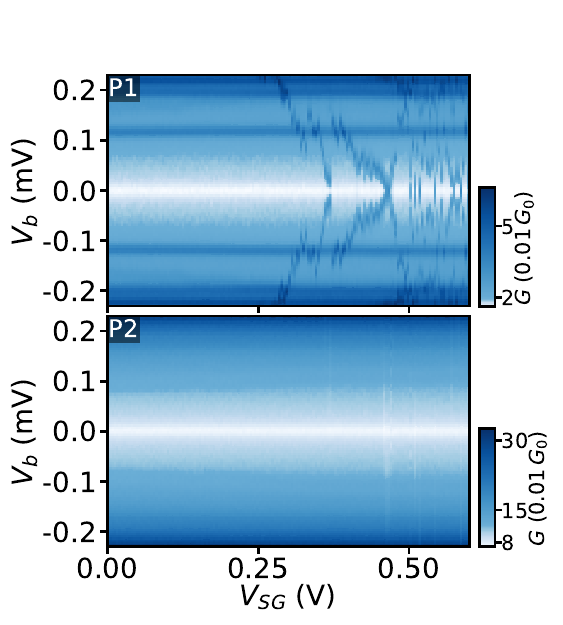}
\caption{\textbf{Effect of the super gate on the tunneling spectroscopy by probes P1 and P2 (Device 1)}: Tunneling conductance $G$ as a function of a bias voltage $V_b$ and a super gate voltage $V_{SG}$ measured by the probes P1 and P2 connected in Setup V2. A parallel magnetic field $B=0.34\,\mathrm{T}$ is applied along the nanowire and the tunnel gate is floated.}\label{fig:d1p1p2sg}
\end{figure}

\begin{figure}[!t] 
\centering
\includegraphics[width=\linewidth] {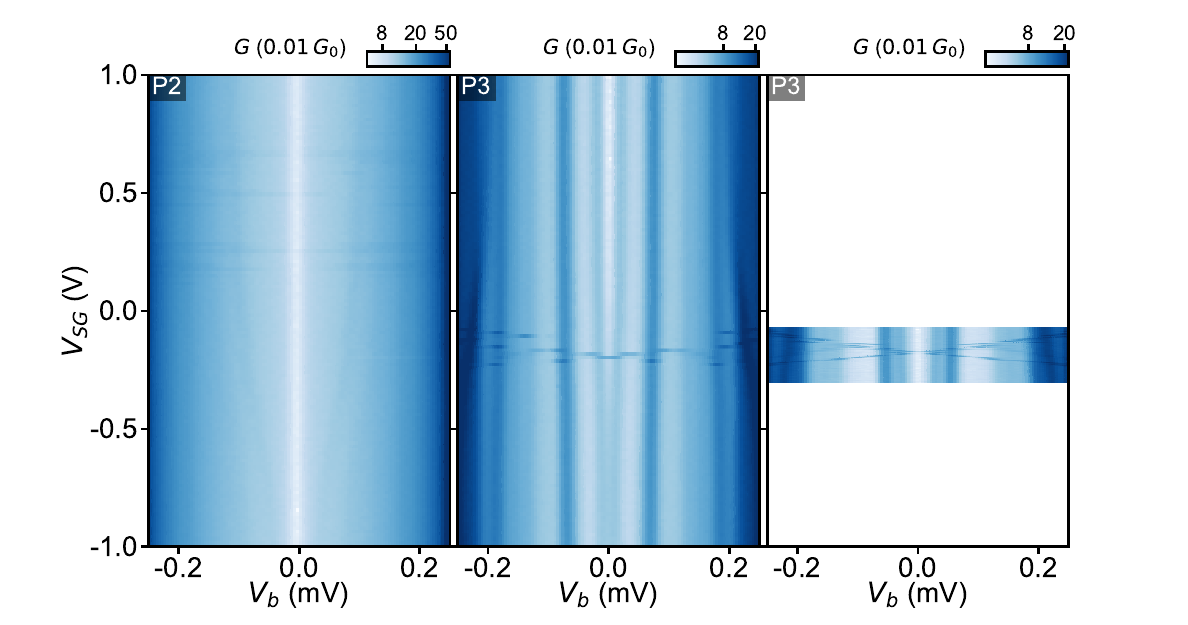}
\caption{\textbf{Effect of the super gate on the tunneling spectroscopy by probes P2 and P3 (Device 1)}: Tunneling conductance $G$ as a function of a bias voltage $V_b$ and a super gate voltage $V_{SG}$ of measured by the probes P2 (left) and P3 (middle) connected in Setup V2. A parallel magnetic field of $B=0.5\,\mathrm{T}$ is applied along the nanowire and the tunnel gate is floated. A narrow $V_{SG}$ range is remeasured by probe P3 in higher resolution (right).}\label{fig:d1p2p3sg}
\end{figure}

\begin{figure}[!t] 
\centering
\includegraphics[width=\linewidth] {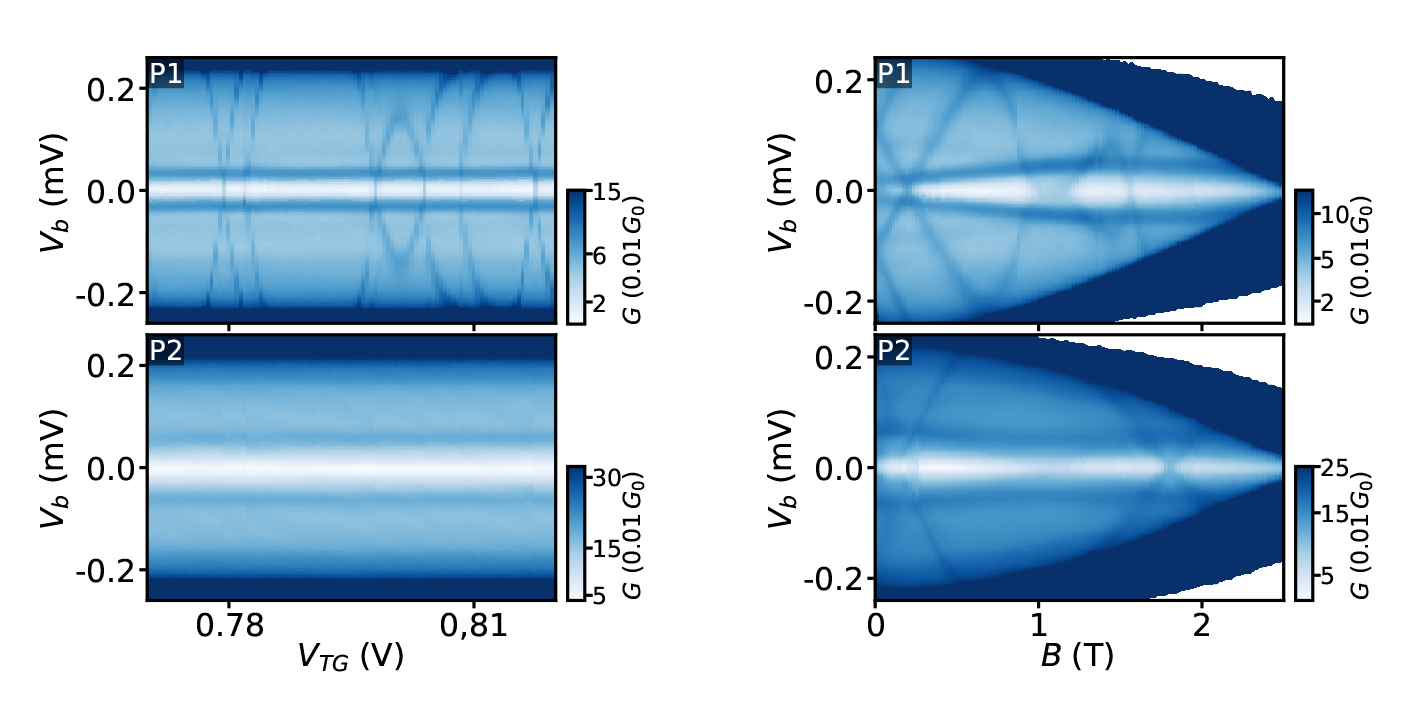}
\caption{\textbf{Effect of the tunnel gate  and parallel magnetic field on the tunneling spectroscopy by probes P1 and P2 (Device 2)}: Tunneling conductance $G$ as a function of a bias voltage $V_b$ and a tunnel gate voltage $V_{TG}$ (left) or a parallel magnetic field $B$ (right) at $V_{SG}=0\,\mathrm{V}$. Different localized subgap states with high $g$ factor are detected by probes P1 and P2. A subgap state tunable by the tunnel gate is detected by P1 and is not detected by P2. These measurements provide evidence of subgap states localized over less than $\sim 200\,\mathrm{nm}$ along the hybrid of Device 2.}\label{fig:d2p1p2tg}
\end{figure}

\begin{figure}[!t] 
\centering
\includegraphics[width=\linewidth] {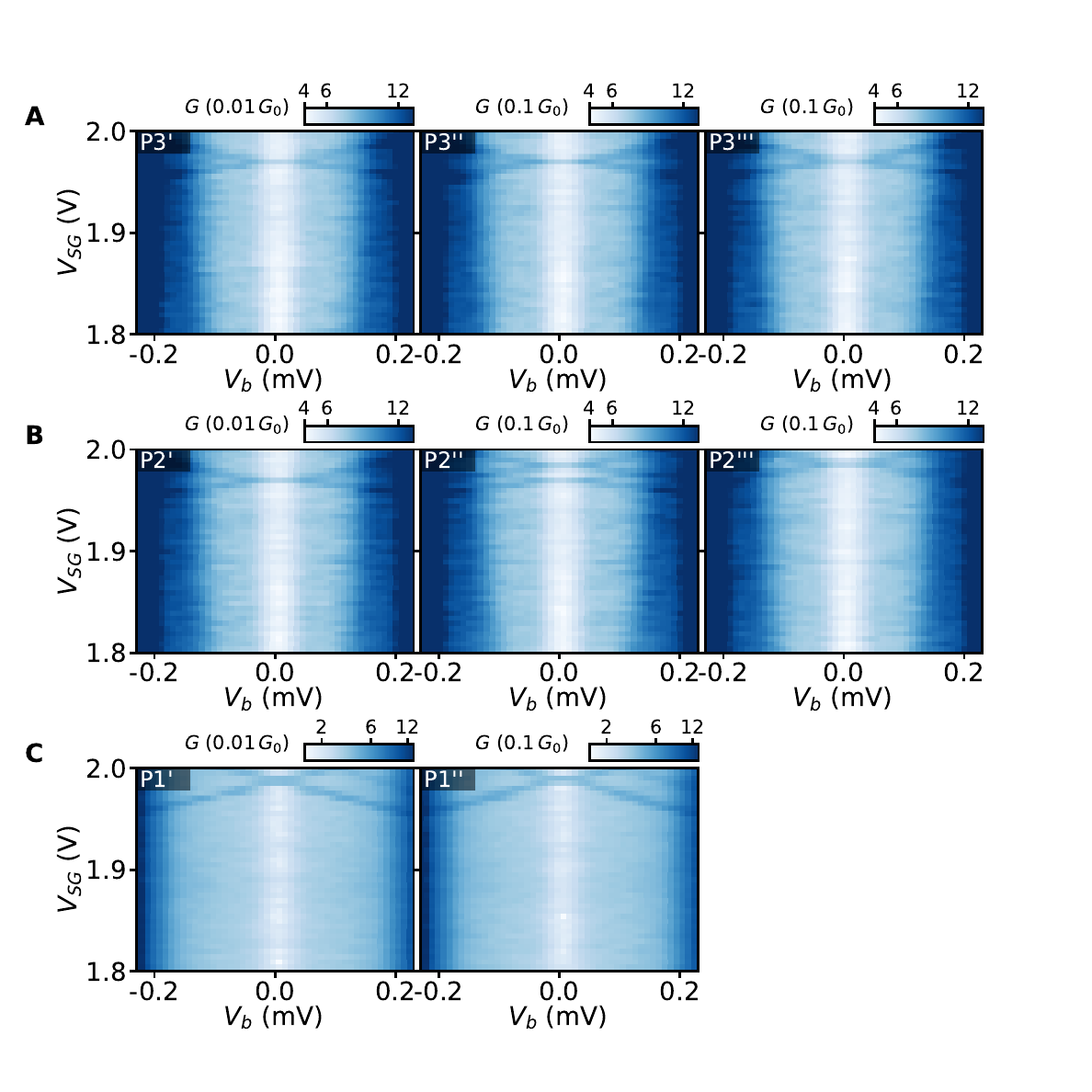}
\caption{\textbf{Detecting a subgap state by multiple probes (Device 3)}: Tunneling conductance $G$ as a function of a bias voltage $V_b$ and a super gate voltage $V_{SG}$ measured by: (\textbf{A}) probe P3, (\textbf{B}) probe P2 and (\textbf{C}) probe P1 connected in Setup V1. Parallel magnetic field is $B=1\,\mathrm{T}$. The probes are in turn connected in the order P3, P2, P1 and the measurement by each probe is repeated multiple times in order to check the charge stability while sweeping $V_{SG}$ (repeated measurements are shown in the same row of (A-C). A single subgap state is detected by both P3 and P2. This state is sensitive to a charge jump observed in the second measurement by P2. A subgap state in the same $V_{SG}$ range is detected by P1. However, the super gate lever arm in this measurement is different – meaning that the subgap state detected by P1 may be different from the one detected by the other two probes.}\label{fig:d3p1p2p3}
\end{figure}

\begin{figure}[!t] 
\centering
\includegraphics[width=\linewidth] {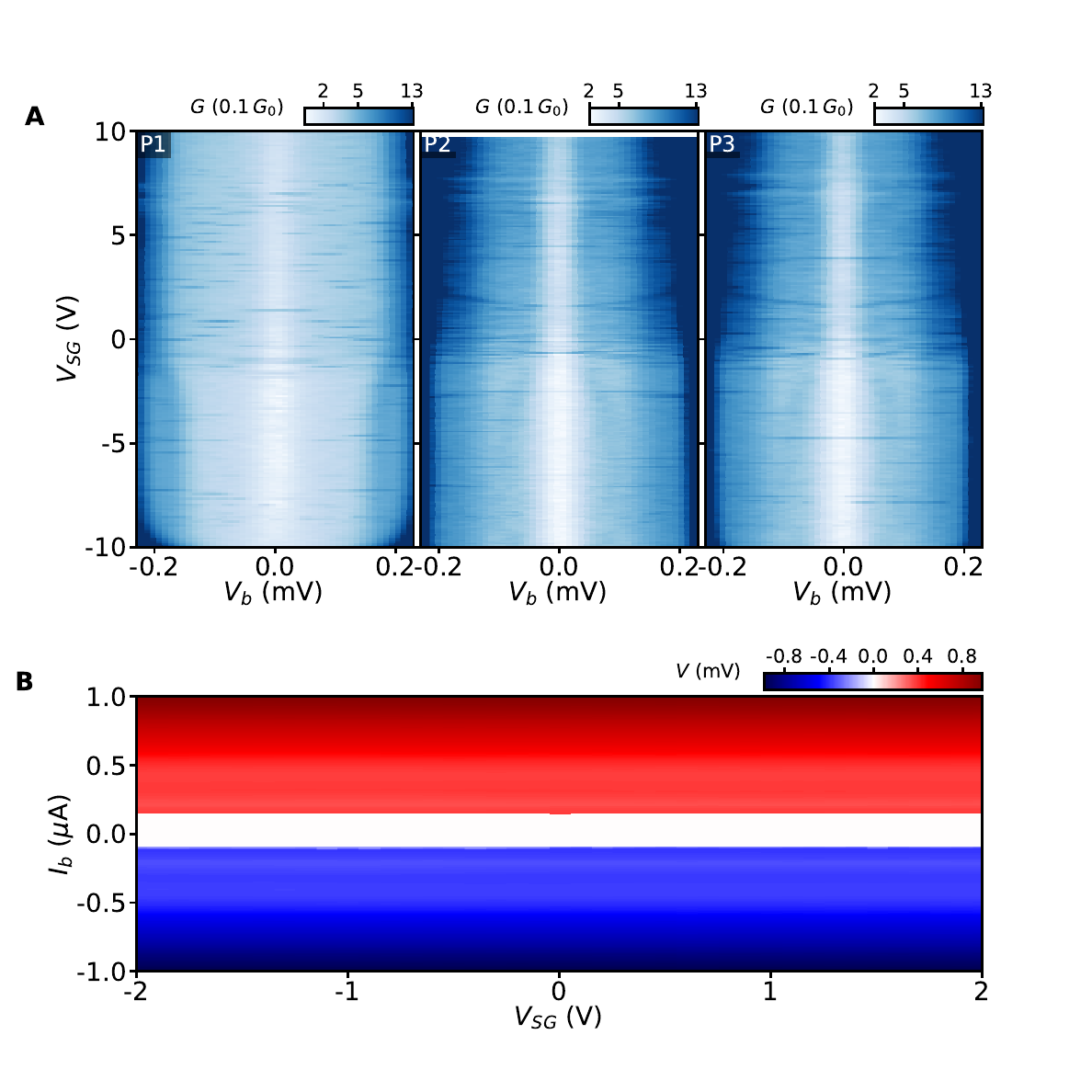}
\caption{\textbf{Effects of sweeping the super gate over broad ranges (Device 3)}: (\textbf{a}) Tunneling conductance $G$ as a function of a bias voltage $V_b$ and a super gate voltage $V_{SG}$ of measured by the probes P1, P2 and P3 in-turn connected in Setup V1. A parallel magnetic field $B=1\,\mathrm{T}$ is applied along the nanowire and the tunnel gate is floated. (\textbf{B}) Voltage drop $V$ as a function of an applied bias current $I_b$ and a super gate voltage $V_{SG}$ at zero field. Probes P1 and P2 are connected in Setup I2 and probe P1 is current-biased. }\label{fig:d3p1p2p3ib}
\end{figure}

\clearpage